\documentclass[11pt,a4paper,twoside]{article}
\newif\ifdraft \global\drafttrue
\def\production{\global\draftfalse}
\production
\usepackage[T1]{fontenc}
\usepackage[latin1]{inputenc}
\usepackage{a4wide}
\usepackage{times}
\usepackage{graphicx}
\usepackage{theorem}
\usepackage[colorlinks=true,hyperindex=true]{hyperref}
\usepackage{color}
\usepackage{fancyhdr}
\usepackage{latexsym}
\usepackage{amsmath}
\usepackage{bbm}
\usepackage{amssymb}
\usepackage{amsfonts}
\usepackage{enumerate}
\usepackage{cancel}
\ifdraft
\fi
\newcounter{smallarabics}

\newcounter{smallroman}

\newcommand{\ben}{\begin{enumerate}[{\rm (1)}]}
\newcommand{\een}{\end{enumerate}}

\newtheorem{theoreme}{Theorem}[section]
\newtheorem{proposition}[theoreme]{Proposition}
\newtheorem{lemma}[theoreme]{Lemma}
\newtheorem{corollary}[theoreme]{Corollary}
\theoremstyle{definition}
{\theorembodyfont{\upshape}
\newtheorem{definition}[theoreme]{Definition}
\newtheorem{remark}[theoreme]{Remark}

}

\def\bep{\begin{proposition}}
\def\eep{\end{proposition}}
\def\bel{\begin{lemma}}
\def\eel{\end{lemma}}
\def\bet{\begin{theoreme}}
\def\eet{\end{theoreme}}
\def\bed{\begin{definition}}
\def\eed{\end{definition}}
\def\bec{\begin{corollary}}
\def\eec{\end{corollary}}
\def\ber{\begin{remark}}
\def\eer{\end{remark}}

\def\rr{{\mathbb R}}

\def\cc{{\mathbb C}}
\def\nn{{\mathbb N}}

\def\textsl{{}}

\def\Im{{\rm Im}\,}
\def\Re{{\rm Re}\,}

\newcommand{\wlim}{\mathop{\mathrm{w-lim}}\limits}

\def\c0inf{C_0^\infty}

\def\proof{\noindent {\bf Proof.}\ \ }

\def\cH{{\cal  H}}

\def\cR{{\cal R}}
\def\cA{{\cal A}}

\def\cB{{\cal B}}

\def\i{{\rm i}}

\newcommand{\beq}{\begin{equation}}
\newcommand{\eeq}{\end{equation}}
\newcommand{\bear}[1]{\begin{array}{#1}}
\newcommand{\ear}{\end{array}}

\def\sp{{\hat e}}

\newcommand{\e}{\mathrm{e}}
\renewcommand{\i}{\mathrm{i}}

\renewcommand{\d}{\mathrm{d}}

\newcommand{\norme}[1]{\|#1\|}

\def\qed{$\Box$\medskip}
\def\cP{{\cal P}}
\def\cC{{\cal C}}

\def\cF{{\cal F}}
\def\cG{{\cal G}}

\def\cN{{\cal N}}

\def\cA{{\cal A}}
\def\cQ{{\cal Q}}

\def\bar{\overline}

\def\12{\frac{1}{2}}

\def\e{{\rm e}}

\def\d{{\rm d}}

\def\one{{\mathbbm 1}}

\def\cH{{\cal H}}

\def\Dom{{\rm Dom}\,}

\def\sp{{\rm sp}}

\def\cS{{\cal S}}
\def\cR{{\cal R}}

\def\fM{\mathfrak{M}}
\def\fH{\mathfrak{H}}
\def\fS{\mathfrak{S}}

\def\tr{{\rm tr}}

\def\P{\mathbb P}
\def\cO{{\cal O}}

\begin{document}
\def\today{}
\title{Energy conservation, counting statistics,\\ and return to equilibrium}
\author{V. Jak\v{s}i\'c$^{1}$,  J. Panangaden$^{1}$, 
A. Panati$^{1,2}$,  C-A. Pillet$^{2}$
\\ \\ 
$^1$Department of Mathematics and Statistics, 
McGill University, \\
805 Sherbrooke Street West, 
Montreal,  QC,  H3A 2K6, Canada
\\ \\
$^2$Aix-Marseille Universit\'e, CNRS, CPT, UMR 7332, Case 907, 13288 Marseille, France\\
Universit\'e de Toulon, CNRS, CPT, UMR 7332, 83957 La Garde, France\\
FRUMAM
}

\maketitle
{\small
{\bf Abstract.} We study a microscopic Hamiltonian model describing an $N$-level 
quantum system $\cS$ coupled to  an infinitely extended thermal reservoir $\cR$. 
Initially, the system $\cS$ is in an arbitrary state while the reservoir is in thermal 
equilibrium at inverse temperature $\beta$. Assuming that the coupled system $\cS+\cR$ 
is mixing with respect to the joint thermal equilibrium state, we study the Full
Counting Statistics (FCS) of the energy transfers $\cS\to\cR$ and $\cR\to\cS$ in the 
process of return to equilibrium. The first FCS describes the increase of the energy
of the system $\cS$. It is an atomic probability measure, denoted
$\P_{\cS,\lambda,t}$,  concentrated on the set of energy differences 
$\sp(H_\cS)-\sp(H_\cS)$ ($H_\cS$ is the Hamiltonian of $\cS$, $t$ is the length of the 
time interval during which the measurement of the energy transfer is performed, and 
$\lambda$ is the strength of the interaction between $\cS$ and $\cR$). The second FCS, 
$\P_{\cR,\lambda,t}$, describes the decrease of the energy of the reservoir $\cR$ and 
is typically a continuous probability measure whose 
support is the whole real line. We study the large time limit $t\rightarrow \infty$ of 
these two measures followed by the weak coupling limit $\lambda \rightarrow 0$ and 
prove that the limiting measures coincide. This result strengthens the first law of 
thermodynamics for open quantum systems. The proofs  are based on modular theory of 
operator algebras and on a representation of $\P_{\cR,\lambda,t}$ by quantum transfer operators.  
 
\thispagestyle{empty}

\section{Introduction} 

The $0^{\rm th}$-law of thermodynamics asserts that a large system, left alone and  
under normal conditions, approaches an equilibrium state characterized by a few 
macroscopic parameters such as temperature and density (see \cite{Ca} or any book on
 thermodynamics). In particular, a small system coupled to a 
large (i.e., infinitely extended) reservoir at temperature $T$ is expected to reach 
its equilibrium state at the same temperature, irrespective of its initial state. 
This specific part of the $0^{\rm th}$-law is often called \emph{return to 
equilibrium}. From a mechanical point of view, return to equilibrium holds if the 
interaction is sufficiently dispersive, which translates into ergodic properties
of the dynamics.

In this paper we consider a quantum system consisting of an $N$-level system $\cS$
coupled to a reservoir $\cR$. We assume that the joint system $\cS+\cR$ has the
property of return to equilibrium. The precise mathematical formulation  of this 
property is given in Section~\ref{sec math setting} (Assumption~\textbf{(M)}).
Although notoriously difficult to prove, return to equilibrium has been 
established for several physically relevant models (spin-boson model, 
spin-fermion model, electronic black box model, locally interacting fermionic systems)  
which are discussed in \cite{AMa,AJPP1,AJPP2,BFS,BM,dRK,DJ,FMU,FMSU,JOP1,JOP2,JP1,JP5}.

We consider the exchange of energy between the system $\cS$ and the reservoir $\cR$. 
Let $\lambda$ be a parameter describing the strength of the coupling between $\cS$ and
$\cR$ and denote by $\Delta \cQ_\cS(\lambda,t)$ the increase of energy of the 
system $\cS$ and by $\Delta \cQ_\cR(\lambda, t)$ the decrease of energy of the 
reservoir $\cR$ during the time period from $0$ to $t$. Set 
\[
\Delta \cQ_\cS= \lim_{\lambda \rightarrow 0} \lim_{t\rightarrow\infty}\Delta \cQ_\cS(\lambda, t), \qquad 
\Delta\cQ_\cR=\lim_{\lambda \rightarrow 0} \lim_{t\rightarrow \infty}\Delta \cQ_\cR(\lambda, t). 
\]
As a consequence of energy conservation, we expect that
\beq
\Delta \cQ_\cS=\Delta \cQ_\cR.
\label{introconservation}
\eeq
This is well understood and can be proven using Araki's perturbation theory, see  
Section~\ref{sec-1st law average}.

The main goal of this paper is to  refine the above result  by considering the Full 
Counting Statistics (abbreviated FCS)
of the energy transfer between ${\cal S}$ and ${\cal R}$. There has been much interest 
in Full Counting Statistics since the seminal paper \cite{LL}.  We refer the reader to
\cite{JOPP}  for   additional information and references to the vast literature on the 
FCS in quantum statistical mechanics.

We consider the probability distribution $\mathbb{P}_{\cS,\lambda,t}$ of the measured 
increase of energy of the system $\cS$ obtained by performing a first measurement at 
time $0$ and a second one at time $t$. While it is straightforward to define  
$\mathbb{P}_{\cS,\lambda,t}$, the analogous definition of the probability measure  
$\mathbb{P}_{\cR,\lambda,t}$ for the reservoir must be given in terms of a relative 
modular operator due to the fact that the reservoir is infinitely extended
(Definition~\ref{defPR}). We devote Section~\ref{refPR} to motivating this definition 
by showing that the correct physical interpretation is recovered in the case of a 
confined reservoir.

Our main result (Theorem~\ref{thm-instant-1}) is the following: under suitable (and in
a certain sense minimal) regularity assumptions the weak limit
\[
\mathbb{P}_\cR=\lim_{\lambda\to0}\lim_{t\to\infty}\P_{\cR,\lambda,t} 
\]
exists and 
\[
\mathbb{P}_\cS=\lim_{\lambda\to0}\lim_{t\to\infty}\P_{\cS,\lambda,t}
=\mathbb{P}_\cR.
\]
Noting that  $\Delta \cQ_\cS/\Delta \cQ_\cR$ is  the first moment of 
$\mathbb{P}_\cS/\mathbb{P}_\cR$, Theorem~\ref{thm-instant-1} is a somewhat surprising 
strengthening of the energy conservation law~\eqref{introconservation}.

Non-equilibrium open quantum systems describing a finite-level system ${\cal S}$ 
coupled to several independent reservoirs in thermal equilibrium at distinct 
temperatures  have different physics and are characterized by steady state energy 
fluxes across $\cS$. An extension of Theorem~\ref{thm-instant-1} to such situations 
requires an approach that differs both technically and conceptually  and is  discussed in the forthcoming paper \cite{JPPP}.

The paper is organized as follows. In Section~\ref{sec math setting} we outline the 
mathematical framework used in this paper and state our assumptions. A review of the 
conservation law~\eqref{introconservation} is given in 
Section~\ref{sec-1st law average}. In Section~\ref{sec main result} we define the FCS
of the energy transfers for the system and reservoir, $\P_{\cS,\lambda,t}$ and 
$\P_{\cR,\lambda,t}$, and state our main result. The definition of 
$\P_{\cR,\lambda,t} $ is motivated in Section~\ref{refPR}. 
Section~\ref{sec-proof-1} is devoted to the proofs.

Throughout the paper we use the algebraic formalism of quantum statistical mechanics 
which is conceptually and technically adapted for the problem under study. The Full 
Counting Statistics has attracted considerable attention in recent experimental and 
theoretical physics literature, and we have attempted to make the paper accessible to 
readers interested in FCS but not familiar with the algebraic formalism. For this 
reason  we give detailed proofs and  collect some well known constructions and results in Appendix~\ref{algebraic}.
This material is standard and can be found in \cite{BR1,BR2}. Modern expositions 
of the algebraic formalism can also be found in \cite{Pi,DJP} and pedagogical 
expositions in \cite{JOPP,Th}. In particular, the exposition in \cite{JOPP} is geared 
toward the application of this algebraic framework to the study of FCS.

\medskip
{\noindent\bf Acknowledgment.} The research of V.J. and J.P. was partly supported by 
NSERC. The research of A.P. was partly supported by NSERC and ANR (grant 12-JS01-0008-01).

\section{Mathematical setting and assumptions}
\label{sec math setting}

We consider a small system $\cS$ coupled to an infinitely extended reservoir $\cR$.

The system $\cS$ is completely determined by a finite dimensional Hilbert space 
$\cH_\cS$, a Hamiltonian $H_\cS$ (a self-adjoint operator on $\cH_\cS$) and a
density matrix $\rho_\cS$ (a positive operator on $\cH_\cS$ 
such that $\tr(\rho_\cS)=1$).

The associated dynamical system is $(\cO_\cS,\tau_\cS,\rho_\cS)$,
where\footnote{Throughout the paper $\cB(\cH)$ denotes the 
set of all bounded operators on a Hilbert space $\cH$.}
$\cO_\cS=\cB(\cH_\cS)$ is the $C^\ast$-algebra of observables of $\cS$,
\[
\tau_\cS^t(A)=\e^{\i tH_\cS}A\e^{-\i tH_\cS}
\]
is the Heisenberg dynamics on ${\cal O}_\cS$ induced by the Hamiltonian $H_\cS$, 
and\footnote{If $\dim \cH<\infty$, we shall identify 
positive linear functionals on $\cB(\cH)$ and positive elements of $\cB(\cH)$ 
according to $\zeta(A)=\tr (\zeta A)$.} $\rho_\cS(A)=\tr(\rho_\cS A)$ is the initial 
state of $\cS$. We do not require $\rho_\cS$ to be faithful. We denote by $\delta_\cS(\,\cdot\,)=\i[H_\cS,\,\cdot\,]$ the generator of $\tau_\cS$.

The reservoir $\cR$ is described by a $C^\ast$-dynamical system 
$(\cO_\cR,\tau_\cR,\omega_\cR)$ in thermal equilibrium at inverse temperature 
$\beta>0$. $\cO_\cR$ is the $C^\ast$-algebra of observables of $\cR$ and 
$\rr\ni t\mapsto\tau_\cR^t$ a strongly continuous group of $\ast$-automorphisms of
$\cO_\cR$ describing the time evolution of $\cR$ in the Heisenberg picture. 
We denote by $\delta_\cR$ its generator, $\tau_\cR^t=\e^{t\delta_\cR}$. Finally,
$\omega_\cR$ is a $(\tau_\cR,\beta)$-KMS state on $\cO_\cR$.

The uncoupled joint system $\cS+\cR$ is described by the $C^\ast$-dynamical 
system $(\cO,\tau_0,\omega)$ where
\[
\cO=\cO_\cS\otimes\cO_\cR,\quad
\tau_0=\tau_\cS\otimes\tau_\cR,\quad
\omega=\rho_\cS\otimes\omega_\cR.
\]
We also introduce the $(\tau_0,\beta)$-KMS state
\[
\omega_0=\omega_\cS\otimes \omega_\cR,
\]
where 
\[
\omega_\cS=\frac{\e^{-\beta H_\cS}}{\tr(\e^{-\beta H_\cS})}
\]
is the thermal equilibrium state of $\cS$.
In the sequel, when tensoring with the identity and  whenever the meaning is clear 
within the context we will omit the identity part. With this convention, the generator 
of $\tau_0$ is given by $\delta_0=\delta_\cS +\delta_\cR$.

Let 
\[
\delta_\lambda=\delta_0+\i\lambda [V,\,\cdot\,],
\]
where $V$ is  a self-adjoint element of $\cO$ and $\lambda$ is a real coupling constant. The interacting dynamics is 
$\tau_\lambda^t=\e^{t\delta_\lambda}$ and the coupled joint system is described by the $C^\ast$-dynamical system
$(\cO,\tau_\lambda,\omega)$.

\ber
With only minor changes all our results and proofs extend to
cases where the reservoir $\cR$ is described by a $W^\ast$-dynamical system 
$(\cO_\cR,\tau_\cR,\omega_\cR)$, and its coupling to $\cS$ by an unbounded 
perturbation $V$ satisfying the general assumptions of \cite{DJP}. The details of this 
generalization can be found in \cite{Pa}.
\eer

To give a precise formulation of the property of return to equilibrium, we recall the  
following result from Araki's perturbation theory of the KMS-structure (see for example 
\cite{BR2,DJP}). Let $(\fH,\pi,\Omega_0)$ be a GNS representation of $\cO$ induced
by the state $\omega_0$. A positive linear functional $\zeta$ on $\cO$ is 
$\omega_0$-normal if it is given by
$$
\zeta(A)=\tr(\rho_\zeta\pi(A))
$$
for some positive trace class operator $\rho_\zeta$ on $\fH$ (see 
Definition~\ref{omeganormdef}). We denote by $\cN$ be the set of all 
$\omega_0$-normal positive linear functionals on $\cO$.
\bet 
\ben
\item There exists a unique $(\tau_\lambda,\beta)$-KMS state in $\cN$ which we denote
by $\omega_\lambda$.
\item The set of all $\omega_\lambda$-normal positive linear functionals on $\cO$
coincides with $\cN$. 
\item
\beq
\lim_{\lambda\to0}\omega_\lambda=\omega_0
\label{araki-lim}
\eeq
holds in the norm topology of the dual space of $\cO$.
\een
\eet

\bed\label{mixing}
The $C^\ast$-dynamical system $(\cO,\tau_\lambda,\omega_\lambda)$ is called 
\emph{ergodic} if, for all states $\zeta\in\cN$ and all $A\in\cO$,
\beq
\lim_{t\to\infty}\frac{1}{t}\int_0^t \zeta(\tau_\lambda^s(A))\d s
=\omega_\lambda(A).
\eeq
It is \emph{mixing} if 
\beq 
\lim_{t\to\infty}\zeta(\tau_\lambda^t(A))=\omega_\lambda(A).
\eeq
\eed
For obvious reasons, ergodicity/mixing of $({\cal O}, \tau_\lambda , \omega_\lambda)$ 
is often called {\em the property of return to equilibrium}. (See \cite{Rob} for 
foundational work on the subject and \cite{BFS,dRK,DJ,FM,JP1,JP2,JP4} for references and additional 
information.)

Our main  dynamical assumption is:

\begin{quote} {\bf Assumption (M)} There exists  $\lambda_0>0$ such that the
$C^\ast$-dynamical system $(\cO,\tau_\lambda,\omega_\lambda)$ is mixing for 
$0<|\lambda|<\lambda_0$.
\end{quote}

We shall also  need the following regularity assumption:

\begin{quote} {\bf Assumption (A)} $V\in\Dom(\delta_\cR)$. 
\end{quote}

\ber\label{confind-rem}
For motivation and clarification purposes we shall 
sometimes consider a {\em confined} reservoir described by a finite dimensional 
Hilbert space $\cH_\cR$ and a Hamiltonian $H_\cR$. In this case $\cO_\cR=\cB(\cH_\cR)$, 
$\delta_\cR(\,\cdot\,)=\i[H_\cR,\,\cdot\,]$, and $\omega_\cR(A)=\tr(\omega_\cR A)$ 
where 
\[
\omega_\cR=\frac{\e^{-\beta H_\cR}}{\tr(\e^{-\beta H_\cR})}.
\]
Moreover, $\cO=\cB(\cH)$ with $\cH=\cH_\cS\otimes\cH_\cR$,
$\tau_\lambda^t(A)=\e^{\i tH_\lambda}A\e^{-\i tH_\lambda}$ with $H_\lambda=H_\cS+H_\cR+\lambda V$, and the $(\tau_\lambda,\beta)$-KMS state
$\omega_\lambda$ is given by the density matrix
$$
\omega_\lambda=\frac{\e^{-\beta H_\lambda}}{\tr(\e^{-\beta H_\lambda})}.
$$
The GNS representation $(\fH,\pi,\Omega_0)$ can be realized in the following way
(see \cite{JOPP}). The Hilbert space $\fH$ is $\cO$ equipped with the inner product 
$\langle X|Y\rangle=\tr(X^\ast Y)$. For $A\in\cO$ the map $\pi(A)\in\cB(\fH)$ is given
by
$$
\pi(A)X=AX.
$$
Finally, the cyclic vector is $\Omega_0=\omega_0^{1/2}$. More generally, any positive
linear functional on $\cO$ can be written as
$$
A\mapsto\tr(\zeta A)=\langle\zeta^{1/2}|\pi(A)\zeta^{1/2}\rangle,
$$
where $\zeta$ is a positive element of ${\cal O}={\mathfrak H}$.
\eer

\section{Review of the first law of thermodynamics}
\label{sec-1st law average}

If Assumption {\bf (M)} holds, then
$$
\lim_{t\to\infty}\omega(\tau_\lambda^t(A))=\omega_\lambda(A)
$$
holds for all $A\in\cO$. We will study the  energy transfer between $\cS$ and $\cR$ 
during the state transition $\omega\to\omega_\lambda$. The energy increase of $\cS$ 
over the time interval $[0, t]$ is 
\beq
\Delta\cQ_\cS(\lambda,t)=\omega(\tau_\lambda^t(H_\cS))-\omega(H_\cS)=
\int_0^t\omega(\tau_\lambda^s(\Phi_\cS))\d s,
\label{DeltaQForm}
\eeq
where
\[
\Phi_\cS=\delta_\lambda(H_\cS)=-\lambda\delta_\cS(V)
\]
is the observable describing the energy flux toward $\cS$. For $0<|\lambda|<\lambda_0$,
the mixing property yields
\[
\Delta\cQ_\cS(\lambda)=\lim_{t\to\infty}\Delta\cQ_\cS(\lambda,t)
=\omega_\lambda(H_\cS)-\omega(H_\cS),
\]
and Eq.~\eqref{araki-lim} further gives
\[
\Delta\cQ_\cS=\lim_{\lambda\to0}\Delta\cQ_\cS(\lambda)
=\omega_\cS(H_\cS)-\rho_\cS(H_\cS).
\]

In what follows we assume that Assumption {\bf (A)} holds. The observable describing 
the  energy flux out of $\cR$ is 
\[
\Phi_\cR=\lambda\delta_\cR(V),
\]
and the decrease of energy of $\cR$ over the time interval $[0,t]$ is 
\beq
\Delta\cQ_\cR(\lambda,t)=\int_0^t\omega(\tau_\lambda^s(\Phi_\cR))\d s.
\label{DeltaQRForm}
\eeq
To motivate this definition, consider a confined reservoir $\cR$. In this case,
according to Remark~\ref{confind-rem}, the decrease of the energy
of $\cR$ is given by
\[
\omega(H_\cR)-\omega(\tau_\lambda^t(H_\cR))
=\int_0^t\omega(\tau_\lambda^s(\Phi_\cR))\d s,
\]
with
$$
\Phi_\cR=-\delta_\lambda(H_\cR)=-\i[H_\lambda,H_\cR]
=\i[H_\cR,\lambda V]=\lambda\delta_\cR(V).
$$

Returning to the general case, since $\Phi_\cR-\Phi_\cS=\lambda\delta_\lambda(V)$, we have 
\beq
\Delta\cQ_\cR(\lambda,t)=\Delta\cQ_\cS(\lambda,t)
+\lambda\omega(\tau_\lambda^t(V)-V).
\label{really}
\eeq
For $0<|\lambda|<\lambda_0$, the mixing property implies 
\[
\Delta\cQ_\cR(\lambda)=\lim_{t\to\infty}\Delta\cQ_\cR(\lambda,t)=
\Delta\cQ_\cS(\lambda)+\lambda(\omega_\lambda(V)-\omega(V)),
\]
and Eq.~\eqref{araki-lim} shows that
\[
\Delta\cQ_\cR=\lim_{\lambda\to0}\Delta\cQ_\cR(\lambda)
\]
satisfies 
\beq 
\Delta\cQ_\cS=\Delta\cQ_\cR.
\label{first-law}
\eeq
Relation~\eqref{first-law} is a mathematical formulation of the first law of
thermodynamics (energy conservation) for the joint system $\cS+\cR$ in the process 
of return to equilibrium described by the above double limit (first $t\to\infty$ and
then $\lambda\to0$).

\section{The first law and full counting statistics}
\label{sec main result}

Our main goal is to refine the previous result by considering the Full Counting 
Statistics of the energy transfer between $\cS$ and $\cR$. We start with the small system $\cS$. Let\footnote{$\sp(A)$ denotes the spectrum of 
the operator $A$.} 
\[
H_\cS=\sum_{e\in\sp(H_\cS)} e P_{e}
\]
be the spectral resolution of $H_\cS$. Suppose that at time $t=0$, when the 
system is in the state $\omega=\rho_\cS\otimes\omega_\cR$, a measurement of $H_\cS$
is performed. The outcome $e$ is observed with probability 
\[
\omega(P_e)=\rho_\cS(P_e).
\]
After the measurement the state of the system is 
\[
\frac{P_e\rho_\cS P_e}{\rho_\cS(P_e)}\otimes\omega_\cR.
\]
This state evolves in time with the dynamics $\tau_\lambda$. 
A second measurement of $H_\cS$ at time $t$ gives $e'$ with probability 
\[
\left(\frac{P_e\rho_\cS P_e}{\rho_\cS(P_e)}
\otimes\omega_\cR\right)(\tau_\lambda^t(P_{e'})).
\]
Hence, $(P_e\rho_\cS P_e\otimes\omega_\cR)(\tau_\lambda^t(P_{e'}))$
is the joint probability distribution of the two measurements. The respective FCS is the atomic probability measure on $\rr$ defined by 
\beq
\P_{\cS,\lambda,t}(S)=\sum_{e'-e\in S}
(P_e\rho_\cS P_e\otimes\omega_\cR)(\tau_\lambda^t(P_{e'})).
\label{work-stat}
\eeq
This measure  is concentrated on the set of energy differences  
$\sp(H_{\cS})-\sp(H_{\cS})$ and $\P_{\cS,\lambda, t}(S)$ is the 
probability  that the measured increase of the energy  of $\cS$ in the above protocol
takes value in the set $S\subset\rr$.  The measure $\P_{\cS, \lambda, t}$ contains full information about the statistics of energy transfer to  ${\cal S}$ over the 
time period $[0, t]$. 

We denote 
by $\langle\, \cdot\,\rangle_{\cS, \lambda, t}$ the expectation with respect to $\P_{\cS, \lambda, t}$ (and similarly 
for other measures that will appear later).
If $\rho_\cS$ and $H_\cS$ commute, then an elementary computation gives 
\[
\langle \varsigma \rangle_{\cS, \lambda, t}=\int_\rr \varsigma \d {\mathbb P}_{\cS, \lambda, t}(\varsigma)=\Delta {\cal Q}_\cS(\lambda, t).
\]
We are interested in the  limiting values of ${\mathbb P}_{\cS, \lambda, t}$  as $t\rightarrow\infty$ and $\lambda \rightarrow 0$.
Assumption {\bf (M)} implies that for $0 <|\lambda|<\lambda_0$, 
\beq
\P_{\cS,\lambda}(S)
=\lim_{t\to\infty}\P_{\cS,\lambda,t}(S)
=\sum_{e'-e\in S}\omega_\lambda(P_{e'})\rho_\cS(P_e).
\label{fri-mor}
\eeq
If, instead of mixing, we assume that $(\cO,\tau_\lambda,\omega_\lambda)$ is ergodic 
for $0 <|\lambda|<\lambda_0$, then~\eqref{fri-mor} holds with 
$\P_{\cS,\lambda,t}$ replaced with 
\[
\frac{1}{t}\int_0^t\P_{\cS,\lambda,s}\d s.
\]

Obviously, 
\[\langle \varsigma\rangle_{\cS, \lambda}=\sum_{e^\prime, e}(e^\prime-e)\omega_\lambda(P_{e'})\rho_\cS(P_e)=
\omega_\lambda(H_\cS)-\rho_\cS(H_\cS)=\Delta \cQ_\cS(\lambda).
\]
Relation~\eqref{araki-lim} further gives
$$
\P_\cS(S)=\lim_{\lambda\to0}\P_{\cS,\lambda}(S)
=\sum_{e'-e\in S}\omega_\cS(P_{e'})\rho_\cS(P_e).
$$
In particular
\[
\langle\varsigma\rangle_\cS=\Delta\cQ_\cS.
\]
Note that  the limiting FCS $\P_\cS$ is the law of $\varsigma=E'-E$ where $E$ and $E'$ are  
independent random variables such that 
$$
{\rm Prob}[E=e]=\rho_\cS(P_e),\qquad
{\rm Prob}[E'=e']=\omega_\cS(P_{e'}).
$$

For later reference, we note that the characteristic function of $\P_\cS$ is 
\beq
\int_\rr\e^{\i\gamma\varsigma}\d\P_\cS(\varsigma)
=\sum_{e,e'\in\sp(H_\cS)}
\omega_\cS(\e^{\i\gamma e'}P_{e'})\rho_\cS(\e^{-\i \gamma e}P_{e})
=\omega_\cS\left(\e^{\i\gamma H_\cS}\right)
\rho_\cS\left(\e^{-\i\gamma H_\cS}\right).
\label{late-1}
\eeq

We now turn to the Full Counting Statistics for the reservoir $\cR$. Recall that
$(\fH,\pi,\Omega_0)$ denotes a GNS representation of $\cO$ induced by the thermal 
equilibrium state of the decoupled system $\omega_0=\omega_\cS\otimes\omega_\cR$.
Let $\fM=\pi(\cO)^{\prime\prime}$ be the associated enveloping von Neumann algebra 
and $\cP$ the natural cone of the pair $(\fM,\Omega_0)$. We denote by $\Omega$ the 
unique vector representative of the initial state $\omega=\rho_\cS\otimes\omega_\cR$ 
in $\cP$ and by $\Delta_{\zeta|\xi}$ the relative modular operator of the two
positive linear functionals $\zeta,\xi\in\cN$. As usual, we set $\Delta_\zeta=\Delta_{\zeta|\zeta}$. Finally, let $\eta=\one\otimes\omega_\cR$.

The following definition will be motivated in Section~\ref{refPR}.
\bed\label{RFCSDef}
The FCS of the decrease of the energy of $\cR$ is the spectral measure 
$\P_{\cR,\lambda,t}$ of the self-adjoint
operator 
\beq
\frac{1}{\beta}\log\Delta_{\eta\circ\tau_\lambda^{-t}|\eta}
\label{defPR}
\eeq
for the vector $\Omega$.
\eed
We note that since $\log\Delta_\eta\Omega=0$ and (see \cite[Section 3]{JP6})  
$$
\log\Delta_{\eta\circ\tau_\lambda^{-t}|\eta}
=\log\Delta_\eta+\beta\int_0^t\tau_\lambda^s(\Phi_\cR)\d s,
$$
Eq.~\eqref{DeltaQRForm} implies
\beq\label{hola}
\langle\varsigma\rangle_{\cR,\lambda,t}=\Delta\cQ_\cR(\lambda, t).
\eeq

The two measures $\P_{\cS,\lambda,t}$ and $\P_{\cR,\lambda,t}$ are of course very 
different. The first one is supported on the discrete set $\sp(H_{\cS})-\sp(H_{\cS})$
while, for an infinitely extended reservoir, the second one is typically a continuous 
measure whose support is the whole real line. On the other hand, the first law gives 
\[
\lim_{\lambda \to0}\lim_{t\to\infty}\langle\varsigma\rangle_{\cR,\lambda,t}
=\lim_{\lambda\to0}\lim_{t\to\infty}\langle\varsigma\rangle_{\cS,\lambda,t},
\]
and one may ask about the relation between the measures $\P_{\cS, \lambda, t}$ and 
$\P_{\cR, \lambda, t}$ in the double limit $t\to\infty$, $\lambda\to0$. 
Our main result is: 
\bet Suppose that Assumptions {\bf (M)} and {\bf (A)} hold and that $0 <|\lambda|<\lambda_0$.  Then  the weak limits
\[
\P_{\cR,\lambda}=\lim_{t\to\infty}\P_{\cR,\lambda,t},
\]
and 
\[
\P_\cR=\lim_{\lambda \rightarrow 0}\P_{\cR,\lambda},
\]
exist, and 
\[
\P_\cR=\P_\cS.
\]
\label{thm-instant-1}
\eet
\ber
\ben
\item For the definition and basic properties of the weak convergence of 
probability measures we refer the reader to Chapter 1 of \cite{Bi}.
\item The proof of Theorem~\ref{thm-instant-1} gives more information and in 
particular provides a formula for the characteristic function of 
$\P_{\cR,\lambda}$ in terms of the modular data of the model.
\item If instead of mixing  we assume ergodicity of 
$(\cO,\tau_\lambda,\omega_\lambda)$ for $0<|\lambda|<\lambda_0$, then 
Theorem~\ref{thm-instant-1} holds with $\P_{\cR, \lambda,t}$ replaced with 
\[
\frac{1}{t}\int_0^t\P_{\cR,\lambda,s}\d s.
\]
\item If $V$ is analytic for $\tau_\cR$ (see Section~2.5.3 in 
\cite{BR1}, or Section~\ref{sec-KMS}), then the 
proof of Theorem \ref{thm-instant-1} considerably simplifies. Moreover,  one easily establishes that in addition
\beq
\lim_{\lambda\to0}\lim_{t\to\infty}\langle \varsigma^n \rangle_{\cR,\lambda,t}
=\lim_{\lambda\to0}\lim_{t\to\infty}\langle \varsigma^n \rangle_{\cS,\lambda,t}
\label{no-sleep}
\eeq
holds for all integers $n>0$.  Details and additional information can be found in 
\cite{Pa}.
\een
\eer
\section{ Motivation of Definition \ref{RFCSDef} }
\label{refPR}

In order to get a physical interpretation of the measure $\P_{\cR,\lambda,t}$,
let us assume that the reservoir $\cR$ is confined. It follows from Definition~\ref{RFCSDef} that 
the characteristic function of $\P_{\cR,\lambda,t}$ is 
$$
\int_\rr\e^{\i\alpha\varsigma}\d\P_{\cR,\lambda,t}(\varsigma)
=\langle\Omega|\Delta_{\eta\circ\tau_\lambda^{-t}|\eta}^{\i\alpha/\beta}\Omega\rangle.
$$
For a confined system, the relative modular operator of two positive linear functionals
$\zeta,\xi$ acts on the GNS Hilbert space $\fH$ as\footnote{Recall Remark~\ref{confind-rem} and see \cite{JOPP}.}
$$
\Delta_{\zeta|\xi}X=\zeta X\xi^{-1}.
$$

It follows that
$$
\Delta_{\eta\circ\tau_\lambda^{-t}|\eta}^{\i\gamma}X
=\e^{\i tH_\lambda}\eta^{\i\gamma}\e^{-\i tH_\lambda}X\eta^{-\i\gamma}.
$$
Let 
\[
H_\cR=\sum_{\varepsilon\in\sp(H_\cR)}\varepsilon P_\varepsilon
\]
be the spectral resolution of $H_\cR$. One then easily computes
\begin{align*}
\int_\rr\e^{\i\alpha\varsigma}\d\P_{\cR,\lambda,t}(\varsigma)
&=\tr\left(\omega^{1/2}\e^{\i tH_\lambda}\eta^{\i\alpha/\beta}\e^{-\i tH_\lambda}\omega^{1/2}\eta^{-\i\alpha/\beta}\right)\\[2mm]
&=\tr\left(\left(\one\otimes\omega_\cR^{\i\alpha/\beta}\right)
\e^{-\i tH_\lambda}\left(\rho_\cS\otimes\omega_\cR^{1-\i\alpha/\beta}\right)
\e^{\i tH_\lambda}\right)\\[2mm]
&=\sum_{\varepsilon,\varepsilon'\in\sp(H_\cR)}
\e^{\i\alpha(\varepsilon-\varepsilon')}
\tr\left((\one\otimes P_{\varepsilon'})\e^{-\i tH_\lambda}
(\one\otimes P_\varepsilon)(\rho_\cS\otimes\omega_\cR)\e^{\i tH_\lambda}\right)\\[2mm]
&=\sum_{\varepsilon,\varepsilon'\in\sp(H_\cR)}
\e^{\i\alpha(\varepsilon-\varepsilon')}
(\rho_\cS\otimes P_\varepsilon\omega_\cR P_\varepsilon)(\tau_\lambda^t(P_{\varepsilon'}))
\end{align*}
from which we can conclude that
$$
\P_{\cR,\lambda,t}(S)
=\sum_{\varepsilon-\varepsilon'\in S}
(\rho_\cS\otimes P_\varepsilon\omega_\cR P_\varepsilon)(\tau_\lambda^t(P_{\varepsilon'})).
$$
Comparing this to Eq.~\eqref{work-stat} leads to an interpretation of 
$\P_{\cR,\lambda,t}$ analogous to the one of $\P_{\cS,\lambda,t}$ (except that we sum 
over $\varepsilon-\varepsilon'\in S$ instead of $\varepsilon'-\varepsilon\in S$ since 
we are measuring the decrease of the energy of $\cR$).
 
Physically relevant infinitely extended reservoirs $\cR$ can be obtained as a 
thermodynamic limit of confined reservoirs $\cR_n$. In such cases, and under very 
general assumptions, the FCS of the infinitely extended system is the weak limit of the 
FCS of confined systems (see Section~5 in~\cite{JOPP}), i.e.,
$$
\P_{\cR,\lambda,t}=\lim_{n\to\infty}\P_{\cR_n,\lambda,t}.
$$
 
The measure $\P_{\cR,\lambda,t}$ contains full information about the statistics of energy transfers out of $\cR$ over the time interval $[0,t]$.

\section{Proofs}
\label{sec-proof-1} 

\subsection{Notation and preliminaries}
\label{sec-prelim}

According to Remark~\ref{confind-rem}, a GNS representation
of $\cO_\cS$ induced by $\omega_\cS$ is given by $(\fH_\cS,\pi_{\cS},\Omega_{\cS})$
where the Hilbert space $\fH_\cS$ is $\cO_\cS=\cB(\cH_\cS)$ equipped with the inner product 
$\langle X|Y\rangle=\tr(X^\ast Y)$. Given  $A\in\cO_\cS$, the linear map $\pi_\cS(A)\in\cB(\fH_\cS)$ is given by
\[
\pi_{\cS}(A)X=AX,
\]
and $\Omega_\cS=\omega_\cS^{1/2}\in\fH_\cS$.
The corresponding natural cone $\cP_\cS\subset\fH_\cS$ and modular 
conjugation $J_\cS:\fH_\cS\to\fH_\cS$ are  
\[
\cP_\cS=\{X\in\fH_\cS\,|\, X\geq 0\},
\quad
J_\cS X=X^\ast.
\]
Let $L_\cS\in\cB(\fH_\cS)$ be defined by
\[
L_\cS X=[H_\cS, X]=(\pi_\cS(H_\cS)-J_\cS\pi_\cS(H_\cS)J_\cS)X.
\]
One easily checks that for all $t\in\rr$ and $A\in\cO$,
\[
\pi_\cS(\tau_{\cS}^t(A))=\e^{\i t L_{\cS}}\pi_\cS(A)\e^{-\i t L_{\cS}},\qquad \e^{-\i t L_{\cS}}{\cal P}_\cS = {\cal P}_\cS, \qquad L_{\cS}\Omega_{\cS}=0.
\]

The operator $L_{\cS}$ is the standard Liouvillean of the dynamical system
$(\cO_\cS,\tau_\cS,\omega_\cS)$.

Let $(\cH_\cR,\pi_\cR,\Omega_\cR)$ be a GNS representation of $\cO_\cR$ induced by the 
state $\omega_\cR$ and denote by $\fM_\cR=\pi_\cR(\cO_\cR)^{\prime\prime}$ the 
associated enveloping von Neumann algebra. Since $\omega_\cR$ is a 
$(\tau_\cR,\beta)$-KMS state, the cyclic vector $\Omega_\cR$ is separating for $\fM_\cR$.
Let $\cP_\cR$, $J_\cR$, $\Delta_\cR$ be the natural cone, modular conjugation,
and modular operator of the pair $(\fM_\cR,\Omega_\cR)$. As a consequence of 
Tomita-Takesaki theory (see Sections~\ref{modular operator app}
and~\ref{GNS liouvillean app} of the Appendix), the standard 
Liouvillean of $(\cO_\cR,\tau_\cR, \omega_\cR)$, i.e., the unique self-adjoint
operator $L_\cR$ on $\fH_\cR$ such that
\[
\pi_\cR(\tau_\cR^t(A))=\e^{\i tL_\cR}\pi_\cR(A)\e^{-\i tL_\cR},\qquad
\e^{-\i tL_\cR}\cP_\cR=\cP_\cR, \qquad
L_\cR\Omega_\cR=0,
\]
for all $t\in\rr$ and $A\in\cO_\cR$, is related to the modular operator by
\[
L_\cR=-\frac{1}{\beta}\log\Delta_\cR.
\]

Set 
\[
\begin{array}{lll}
\fH=\fH_\cS\otimes\fH_\cR,&\pi=\pi_\cS\otimes \pi_\cR,
&\Omega_0=\Omega_\cS\otimes\Omega_\cR,\\[3mm]
\fM=\pi_\cS(\cO_\cS)\otimes\fM_\cR,&\cP=\cP_\cS\otimes\cP_\cR,&
J=J_\cS\otimes J_\cR,\\[3mm]
&L_0=L_\cS+L_\cR.& \\[3mm]
\end{array}
\]
The triple $(\fH,\pi,\Omega_0)$ is a GNS representation of $\cO$ induced 
by $\omega_0$. $\fM=\pi({\cal O})^{\prime\prime}$ and the  natural cone and modular 
conjugation of the pair $(\fM,\Omega_0)$ are $\cP$ and $J$. The operator $L_0$ is the 
standard Liouvillean of $(\cO,\tau_0,\omega_0)$. Moreover, for any $t\in\rr$ and 
$A\in\cO$, one has
\[
\pi(\tau_\lambda^t(A))=\e^{\i t(L_0+\lambda \pi(V))}\pi(A)\e^{-\i t(L_0+\lambda\pi(V))}.
\]

Recall that a positive linear functional $\zeta$ on $\cO$ belongs to $\cN$ (i.e., is
$\omega_0$-normal) iff there exists a positive trace class operator $\rho_\zeta$ on 
$\fH$ such that, for all $A\in\cO$,
\[
\zeta(A)=\tr(\rho_\zeta\pi(A)).
\]
Such a functional obviously extends to $\fM$, and we denote this extension by the same 
letter. As a consequence of modular theory (see Section~\ref{sec normal states}) there 
exists a unique vector $\Omega_\zeta\in\cP$, called the standard vector representative of $\zeta$,  such that
\[
\zeta(A)=\langle\Omega_\zeta|A \Omega_\zeta\rangle
\]
 for all $A\in\fM$.
The standard vector representative of $\eta=\one\otimes\omega_\cR$ is 
\[
\Omega_\eta=\one\otimes \Omega_\cR,
\] 
and the standard vector representative $\Omega=\Omega_\omega$ of the initial state $\omega=\rho_\cS\otimes \omega_\cR$ is
\beq
\Omega=\rho_\cS^{1/2}\otimes\Omega_\cR=\pi(\rho_\cS^{1/2}\otimes\one)\Omega_\eta.
\label{OmegaForm}
\eeq

For later reference, we recall some results of Araki's perturbation theory of the KMS 
structure (see \cite[Chapter~5]{BR2} and \cite{DJP}). First
\beq
\Omega_0\in\Dom\left(\e^{-\frac{\beta}{2}(L_0+\lambda \pi(V))}\right),
\label{araki-1}
\eeq
and the vector 
\beq\Omega_\lambda =\e^{-\frac{\beta}{2}(L_0+ \lambda \pi(V))}\Omega_0
\eeq
is well-defined. Moreover, the vector-valued function 
\beq
z\mapsto \e^{-z(L_0+\lambda \pi(V))}\Omega_0\in\fH
\label{araki-3}
\eeq
is analytic inside the strip $0<\Re z<\beta/2$, and norm continuous and bounded on its
closure. The map 
\[
\rr\ni\lambda\mapsto\Omega_\lambda\in\fH
\]
is real analytic, $\Omega_\lambda\in\cP$, and, for $A\in\fM$, 
\[
\omega_\lambda(A)=\frac{\langle\Omega_\lambda|A\Omega_\lambda\rangle}
{\|\Omega_\lambda\|^2}.
\]
The standard Liouvillean of $(\cO,\tau_\lambda,\omega_\lambda)$ is given by
\[
L_\lambda=L_0+\lambda\pi(V)-\lambda J\pi(V)J,
\]
and it satisfies 
\[
\pi(\tau_\lambda^t(A))=\e^{\i tL_\lambda}\pi(A)\e^{-\i tL_\lambda},\qquad
\e^{-\i t L_\lambda}\cP=\cP,\qquad
L_\lambda\Omega_\lambda=0.
\]

It is well-known \cite{BR2,JP1,Pi} that the ergodic properties of 
$(\cO,\tau_\lambda,\omega_\lambda)$ can be characterized in terms of the spectral
properties of $L_\lambda$.\footnote{This aspect of modular theory is sometimes called 
Quantum Koopmanism.} More precisely, $(\cO,\tau_\lambda,\omega_\lambda)$ is ergodic 
iff $0$ is a simple eigenvalue of $L_\lambda$ and mixing iff 
\beq
\wlim_{|t|\rightarrow \infty}\e^{\i tL_\lambda}
=\frac{|\Omega_\lambda\rangle\langle\Omega_\lambda|}{\|\Omega_\lambda\|^2}.
\label{to-recall}
\eeq
In particular, the last relation holds if the spectrum of $L_\lambda$ on the orthogonal
complement of $\cc\Omega_\lambda$ is purely absolutely continuous.

Recall that $\eta=\one\otimes \omega_\cR$ with $\omega_\cR$ a $(\tau_\cR,\beta)$-KMS 
state. The proof of Theorem~\ref{thm-instant-1} is centered around the function 
$$
\cF_{\lambda,t}(\alpha)
=\langle\Omega|\Delta_{\eta\circ\tau_\lambda^{-t}|\eta}^\alpha\Omega\rangle
=\int_\rr\e^{\alpha\beta\varsigma}\d\P_{\cR,\lambda,t}(\varsigma).
$$
By the properties of the weak convergence of measures (see Chapter 1 in~\cite{Bi}),  Theorem~\ref{thm-instant-1} is equivalent to the following statements: for 
$0<|\lambda|<\lambda_0$ and $\gamma\in\rr$, the limit 
\beq
\cF_\lambda(\i\gamma/\beta)=\lim_{t\to\infty}\cF_{\lambda,t}(\i\gamma/\beta)
\label{sale-1}
\eeq
exists and defines a continuous function $\rr\ni\gamma\mapsto\cF_\lambda(\i\gamma/\beta)$ such that, for $\gamma\in\rr$, 
\beq
\lim_{\lambda\to0}\cF_\lambda(\i\gamma/\beta)=
\omega_\cS\left(\e^{\i\gamma H_\cS}\right)
\rho_\cS\left(\e^{-\i\gamma H_\cS}\right).
\label{sale-2}
\eeq
By~\eqref{late-1}, the right-hand side in~\eqref{sale-2} is the characteristic function
of $\P_\cS$. The existence of the limit~\eqref{sale-1} and the continuity of
$\gamma\mapsto\cF_\lambda(\i\gamma/\beta)$ are equivalent to the statement that 
$\P_{\cR,\lambda,t}$ converges weakly as $t\rightarrow \infty$ to a probability measure 
$\P_{\cR,\lambda}$ whose characteristic function is $\cF_\lambda(\i \gamma/\beta)$.
The relation~\eqref{sale-2} is equivalent to the statement that $\P_{\cR,\lambda}$ 
converges weakly to $\P_\cS$ as $\lambda \rightarrow 0$.

We finish this section by recalling some basic properties of the relative modular
operator $\Delta_{\eta\circ\tau_\lambda^{-t}|\eta}$. First, since 
$\eta=\one\otimes\omega_\cR\in\cN$, 
the modular conjugation $J_\eta$ and the modular operator $\Delta_\eta$ of the
pair $(\fM,\Omega_\eta)$ are given by $J_\eta=J$, $\Delta_\eta=\e^{-\beta L_\cR}$ (see 
Proposition~\ref{PropJ} and Eq.~\eqref{LDeltaForm}), and 
\beq
J^\ast J=J^2=1,\quad
\Delta_\eta\Omega_\eta=\Omega_\eta,\quad
J\Delta_\eta^{\i s}=\Delta^{\i s}_\eta J.
\label{JDeltaProp}
\eeq
Moreover, by the result of \cite[Eq.~(2.13)]{JP3},
\beq
\Delta_{\eta\circ\tau_\lambda^{-t}|\eta}
=\e^{\i t(L_0+\lambda\pi(V))}\Delta_\eta\e^{-\i t(L_0+\lambda\pi(V))}
=\Gamma_\lambda(t)\Delta_\eta\Gamma_\lambda^\ast(t),
\label{cocyclform}
\eeq
where $\Gamma_\lambda(t)$ is the unitary
\beq
\Gamma_\lambda(t)=\e^{\i t(L_0+\lambda\pi(V))}\e^{-\i tL_0}.
\label{GammaDef}
\eeq
One easily checks that $\Gamma_\lambda(t)$ satisfies the Cauchy problem
$$
\partial_t\Gamma_\lambda(t)=\i\lambda\Gamma_\lambda(t)\pi(\tau_0^t(V)),\quad
\Gamma_\lambda(0)=\one.
$$
Hence, for any $B\in\fM'$,
$$
\partial_t[B,\Gamma_\lambda(t)]=\i\lambda[B,\Gamma_\lambda(t)]\pi(\tau_0^t(V)),\quad
[B,\Gamma_\lambda(0)]=0,
$$
and uniqueness of the solution of the Cauchy problem gives that 
\beq
\Gamma_\lambda(t)\in\fM''=\fM
\label{late}
\eeq
for all $\lambda,t\in\rr$.

\subsection{Proof of Theorem \ref{thm-instant-1}}

We start by establishing a few basic properties of the function
$$
\cF_{\lambda,t}(\alpha)
=\langle\Omega|\Delta_{\eta\circ\tau_\lambda^{-t}|\eta}^\alpha\Omega\rangle
=\|\Delta_{\eta\circ\tau_\lambda^{-t}|\eta}^{\alpha/2}\Omega\|^2
=\int_\rr \e^{\alpha\beta\varsigma}
\d\P_{\cR,\lambda,t}(\varsigma).
$$

\bel\label{rain}
For any $\lambda\in\rr$ and any $t\in\rr$, one has
\[
\cF_{\lambda,t}(0)=1,\qquad
0\le\cF_{\lambda,t}(1)\leq\dim\cH_\cS.
\]
\eel
\proof
The relations $\cF_{\lambda,t}(0)=1$ and $\cF_{\lambda,t}(1)\geq 0$ are obvious.
By Eq.~\eqref{OmegaForm} and~\eqref{cocyclform}, one has
$$
\cF_{\lambda,t}(1)
=\|\Delta_{\eta\circ\tau_\lambda^{-t}|\eta}^\12\Omega\|^2
=\|\Delta_\eta^\12\Gamma_\lambda^\ast(t)\pi(\rho_\cS^\12\otimes\one)\Omega_\eta\|^2.
$$
Using~\eqref{late} and the anti-unitarity of the modular conjugation $J$ we derive
\[
\cF_{\lambda,t}(1)
=\|J\Delta_\eta^\12\Gamma_\lambda^\ast(t)\pi(\rho_\cS^\12\otimes\one)\Omega_\eta\|^2
=\|\pi(\rho_\cS^{\frac{1}{2}}\otimes\one)\Gamma_\lambda(t)\Omega_\eta\|^2\leq \|\Omega_\eta\|^2=\dim\cH_\cS.
\]
\hfill\qed

For $S\subset\rr$, we denote
\[
\fS(S)=\{z\in\cc\,|\,\Re z\in S\}.
\]

\bel\label{rain-1}
For any $\lambda\in\rr$ and any $t\in\rr$,
the function $\alpha\mapsto\cF_{\lambda,t}(\alpha)$ is analytic in the strip 
$\fS(]0,1[)$ and bounded and continuous on its closure. Moreover, the bound
\beq\label{Fbound1}
\sup_{t\in\rr}|\cF_{\lambda,t}(\alpha)|\leq 1+(\dim\cH_\cS-1)\Re\alpha
\eeq
holds for any $\alpha\in\fS([0,1])$.
\eel

\proof For $\alpha\in[0,1]$, the convexity of the exponential function yields
$$
\e^{\alpha\beta\varsigma}\le(1-\alpha)+\alpha\e^{\beta\varsigma}
$$
so that, by the previous Lemma,
$$
\cF_{\lambda,t}(\alpha)\le(1-\alpha)\cF_{\lambda,t}(0)+\alpha\cF_{\lambda,t}(1)
\leq1+(\dim\cH_\cS-1)\alpha.
$$
The fact that $|\e^z|=\e^{\Re z}$ yields the rigidity of $\cF_{\lambda,t}$, i.e.,
$$
|\cF_{\lambda,t}(\alpha)|\le\cF_{\lambda,t}(\Re\alpha),
$$
from which the bound~\eqref{Fbound1} follows. Writing $\cF_{\lambda,t}(\alpha)$ as
the sum of the two Laplace transforms
$$
\cF_{\lambda,t}^\pm(\alpha)
=\int_{\rr^\pm}\e^{\alpha\beta\varsigma}\d\P_{\cR,\lambda,t}(\varsigma),
$$
we deduce from the previous estimate that $\cF_{\lambda,t}^+$ (resp. 
$\cF_{\lambda,t}^-$) is analytic on the strip $\fS(]-\infty,1[)$ 
(resp. $\fS(]0,\infty[)$). Hence, $\cF_{\lambda,t}$ is analytic on $\fS(]0,1[)$.
For $\alpha,\alpha'\in\fS(]-\infty,1])$, one has
$$
|\cF_{\lambda,t}^+(\alpha')-\cF_{\lambda,t}^+(\alpha)|
\le\int_{\rr^+}|\e^{\alpha'\beta\varsigma}-\e^{\alpha\beta\varsigma}|
\d\P_{\cR,\lambda,t}(\varsigma),
$$
and the simple estimate
$$
|\e^{\alpha'\beta\varsigma}-\e^{\alpha\beta\varsigma}|
\le2\,\e^{\beta\varsigma}
$$
allows us to apply the dominated convergence theorem and conclude that
$\cF_{\lambda,t}^+$ is continuous in $\fS(]-\infty,1])$. One shows in a similar way
that $\cF_{\lambda,t}^-$ is continuous in $\fS([0,\infty[)$.
\hfill\qed

\bel\label{rain-2}
For any $\lambda\in\rr$ and any $\delta\in]0,1[$, one has
\[
\sup_{t\in\rr,\alpha\in\fS(]0,\delta[)}|\partial_\alpha\cF_{\lambda,t}(\alpha)|<\infty.
\]
\eel
\proof
By Relation~\eqref{hola}, one has 
$$
\Delta\cQ_\cR(\lambda,t)
=-\int_{\rr^-}|\varsigma|\,\d\P_{\cR,\lambda,t}(\varsigma)
+\int_{\rr^+}\varsigma\,\d\P_{\cR,\lambda,t}(\varsigma).
$$
The inequality $\e^x\geq x$ and Lemma~\ref{rain} further give
$$
\int_{\rr^+}\varsigma\,\d\P_{\cR,\lambda,t}(\varsigma)
\leq\beta^{-1}\int_{\rr^+}\e^{\beta\varsigma}\d\P_{\cR,\lambda,t}(\varsigma)
\leq\beta^{-1}\cF_{\lambda,t}(1)\leq\beta^{-1}\dim\cH_\cS.
$$
It follows that
$$
\int_{\rr^-}|\varsigma|\,\d\P_{\cR,\lambda,t}(\varsigma)
\leq\beta^{-1}\dim\cH_\cS-\Delta\cQ_\cR(\lambda,t).
$$
By Lemma~\ref{rain-1} and a well known property of Laplace transforms one has, for $\alpha\in\fS(]0,1[)$,
$$
\partial_\alpha\cF_{\lambda,t}(\alpha)=\beta\int_\rr\varsigma
\e^{\alpha\beta\varsigma}
\d\P_{\cR,\lambda,t}(\varsigma).
$$
Thus, using the elementary estimate $x\e^{ax}\leq(1-a)^{-1}\e^x$ valid for
$a<1$ and $x\in\rr$, we further get
\begin{align*}
|\partial_\alpha\cF_{\lambda,t}(\alpha)|
&\leq\beta\int_{\rr^-}|\varsigma|\,\d\P_{\cR,\lambda,t}(\varsigma)
+\int_{\rr^+}\beta\varsigma\e^{(\Re\alpha)\beta\varsigma}\d\P_{\cR,\lambda,t}(\varsigma)\\
&\leq\dim\cH_\cS-\beta\Delta\cQ_\cR(\lambda,t)+(1-\Re\alpha)^{-1}\cF_{\lambda,t}(1)\\[8pt]
&\leq(1+(1-\Re\alpha)^{-1})\dim\cH_\cS-\beta\Delta\cQ_\cR(\lambda,t).
\end{align*}
The result now follows from Eq.~\eqref{DeltaQForm} and~\eqref{really} which imply
that $|\Delta\cQ_\cR(\lambda,t)|\leq2\|H_\cS+\lambda V\|$.
\hfill\qed

Next, we derive an alternative representation of the function $\cF_{\lambda,t}$ on the
line $\12+\i\rr$. To this end, we set 
\[
\widehat{L}_\lambda=L_\cR+\pi(H_\cS+\lambda V),\qquad 
\widehat{\Omega}=\pi(\rho_\cS^\12\otimes\one)\Omega. 
\]

\bel\label{prop-instant-1}
For any $\lambda,t,s\in \rr$, one has
\[
\cF_{\lambda,t}\left(\12+\i s\right)
=\langle\e^{\i\beta s\widehat{L}_\lambda}\widehat{\Omega}|\e^{\i tL_\lambda}
\e^{\i\beta s\widehat{L}_\lambda}\Omega_\eta\rangle.
\]
\eel

\proof 
Set $R=\pi(\rho_\cS^\12\otimes\one)\in\fM$. Eq.~\eqref{OmegaForm}, \eqref{cocyclform} and the relation $\Delta_\eta^\12 A\Omega_\eta=JA^\ast\Omega_\eta$ yield
\beq
\Delta_{\eta\circ\tau_\lambda^{-t}|\eta}^{\12+\i s}\Omega
=\Gamma_\lambda(t)\Delta_\eta^{\i s}
\Delta_\eta^\12\Gamma_\lambda^\ast(t)R\Omega_\eta
=\Gamma_\lambda(t)\Delta_\eta^{\i s}JR\Gamma_\lambda(t)\Omega_\eta.
\label{comb-1}
\eeq
Using Eq.~\eqref{JDeltaProp} and the fact that $R$ commutes with 
$\Delta_\eta^{\i s}=\e^{-\i\beta sL_\cR}$, we derive
\beq
\Gamma_\lambda(t)\Delta_\eta^{\i s}JR\Gamma_\lambda(t)\Omega_\eta
=\Gamma_\lambda(t)JR\Delta_\eta^{\i s}\Gamma_\lambda(t)\Omega_\eta
=\Gamma_\lambda(t)(JRJ)J\Delta_\eta^{\i s}\Gamma_\lambda(t)\Omega_\eta.
\eeq
Since $JRJ\in\fM^\prime$, relation~\eqref{late} and the fact that 
$\Delta_\eta^{-\i s}J\Omega_\eta=\Omega_\eta$ yield
\beq
\Gamma_\lambda(t)(JRJ)J\Delta_\eta^{\i s}\Gamma_\lambda(t)\Omega_\eta
=(JRJ)\Gamma_\lambda(t)J\Delta_\eta^{\i s}\Gamma_\lambda(t)J\Delta_\eta^{-\i s}J\Omega_\eta
=(JRJ)\Gamma_\lambda(t)\widetilde{\Gamma}_\lambda(t)\Omega_\eta,
\label{comb-3}
\eeq
where 
\beq
\widetilde{\Gamma}_\lambda(t)
=J\Delta_\eta^{\i s}\Gamma_\lambda(t)\Delta_\eta^{-\i s}J.
\label{GammaTildeDef}
\eeq
Since $JRJ\Omega=\widehat{\Omega}$, it follows from~\eqref{comb-1}--\eqref{comb-3} that
\beq
\cF_{\lambda,t}\left(\12+\i s\right)
=\langle\widehat{\Omega}|
\Gamma_\lambda(t)\widetilde{\Gamma}_\lambda(t)\Omega_\eta\rangle.
\label{calFform1}
\eeq

For $A=A^\ast\in\cB(\fH)$, set $A_t=\e^{\i tL_\cR}A\e^{-\i tL_\cR}$ and denote by
$U_A^t$ is the solution of the Cauchy problem
$$
\partial_t U_A^t=\i U_A^t\,A_t,\quad U_A(0)=\one.
$$
The following properties are easy consequences of the uniqueness of the solution to
this problem.
\ben
\item $U_A^t=\e^{\i t(L_\cR+A)}\e^{-\i tL_\cR}$.
\item If $A\in\fM$, then $U_A^t\in\fM$.
\item If $A,B\in\fM$, then $U_{A-JBJ}^t=U_A^tJU_B^tJ$.
\item $\Delta_\eta^{\i s}U_A^t\Delta_\eta^{-\i s}=U_{A_{-\i\beta s}}^t$
\een
Set $M=\pi(H_\cS+\lambda V)$. Since $L_0=L_\cR+\pi(H_\cS)-J\pi(H_\cS)J$ and
$L_0+\lambda \pi(V)=L_\cR+M-J\pi(H_\cS)J$, it follows 
from~\eqref{GammaDef} and~\eqref{GammaTildeDef} that
$$
\Gamma_\lambda(t)=U_M^tU_{\pi(H_\cS)}^{t\ast},\qquad
\widetilde{\Gamma}_\lambda(t)=(JU_{M_{-\i\beta s}}^tJ)(JU_{\pi(H_\cS)}^tJ)^\ast,
$$
and hence
$$
\Gamma_\lambda(t)\widetilde{\Gamma}_\lambda(t)
=(U_M^tJU_{M_{-\i\beta s}}^tJ) (U_{\pi(H_\cS)}^tJU_{\pi(H_\cS)}^tJ)^\ast
=U_M^tJU_{M_{-\i\beta s}}^tJ\,\e^{\i tL_\cR}\e^{-\i tL_0}.
$$
From the fact that $\widehat{L}_\lambda=L_\cR+M$ and
$L_\lambda=L_\cR+M-JMJ$, we deduce
\begin{align*}
\e^{\i tL_\lambda}\e^{\i\beta s\widehat{L}_\lambda}
&=(U_M^tJU_M^tJ\e^{\i tL_\cR})
U_M^{\beta s}\e^{\i\beta sL_\cR}\\
&=U_M^t(JU_M^tJ)(\e^{\i tL_\cR}U_M^{\beta s}\e^{-\i tL_\cR})
\e^{\i tL_\cR}\e^{\i\beta sL_\cR}\\
&=U_M^t(\e^{\i tL_\cR}U_M^{\beta s}\e^{-\i tL_\cR})\e^{\i\beta sL_\cR}
\e^{-\i\beta sL_\cR}
(JU_M^tJ)\e^{\i\beta sL_\cR}\e^{\i tL_\cR}\\
&=\e^{\i t\widehat{L}_\lambda}\e^{\i\beta s\widehat{L}_\lambda}
\e^{-\i tL_\cR}(JU_{M_{-\i\beta s}}^tJ)\e^{\i tL_\cR}\\
&=\e^{\i\beta s\widehat{L}_\lambda}U_M^tJU_{M_{-\i\beta s}}^tJ\e^{\i tL_\cR}\\
&=\e^{\i\beta s\widehat{L}_\lambda}\Gamma_\lambda(t)
\widetilde{\Gamma}_\lambda(t)\e^{\i tL_0}.
\end{align*}
Since $\e^{\i tL_0}\Omega_\eta=\Omega_\eta$, the result follows from 
Eq.~\eqref{calFform1}.
\hfill\qed

We are now in position to investigate the behavior of $\cF_{\lambda,t}$ in the limits 
$t\to\infty$ and $\lambda\to0$.

\bel Suppose that $0<|\lambda|<\lambda_0$. Then the limit 
\beq
\cF_\lambda(\alpha)=\lim_{t\to\infty}\cF_{\lambda,t}(\alpha)
\label{limit-rain}
\eeq
exists for $\alpha\in\fS([0,1 [)$. The function $\alpha\mapsto\cF_\lambda(\alpha)$ is
analytic on $\fS(]0, 1[)$ and continuous on $\fS([0,1[)$. Moreover,
\[
\cF_\lambda\left(\12+\i s\right)=\frac{1}{\|\Omega_\lambda\|^2}
\langle\widehat{\Omega}|\e^{-\i\beta s\widehat{L}_\lambda}\Omega_\lambda\rangle
\langle\Omega_\lambda|\e^{\i\beta s \widehat{L}_\lambda}\Omega_\eta\rangle
\]
holds for $s\in\rr$. 
\eel
\proof The previous lemma and Eq.~\eqref{to-recall} yield that for 
$0<|\lambda|<\lambda_0$ and $s\in\rr$, 
\[
\lim_{t\to\infty}\cF_{\lambda,t}\left(\12+\i s\right)
=\frac{1}{\|\Omega_\lambda\|^2}
\langle\widehat{\Omega}|\e^{-\i\beta s\widehat{L}_\lambda}\Omega_\lambda\rangle
\langle\Omega_\lambda|\e^{\i\beta s\widehat{L}_\lambda}\Omega_\eta\rangle.
\]
Lemma~\ref{rain-1} and Vitali's convergence theorem imply that the 
limit~\eqref{limit-rain} exists uniformly on compact subsets of 
$\fS(]0,1[)$, and that the limiting function 
$\alpha\mapsto\cF_\lambda(\alpha)$ is analytic on $\fS(]0,1[)$.

Let $K\subset\fS([0,1[)$ be compact. Then there exist $\delta\in]0,1[$ and $k>0$
such that $K\subset\widehat{K}=[0,\delta]+\i[-k,k]$. Set
$$
C_\delta
=\sup_{t\in\rr,\alpha\in\fS(]0,\delta[)}|\partial_\alpha\cF_{\lambda,t}(\alpha)|.
$$
By Lemma~\ref{rain-2}, $C_\delta <\infty$.  If $\epsilon>0$ 
is small enough, then $r=\epsilon/12C_\delta<\delta$ and the uniform convergence on compacts in~\eqref{limit-rain}
ensures that there exists $T>0$ such that $|\cF_{\lambda,t}(\alpha)-\cF_{\lambda}(\alpha)|<\epsilon/3$ for any
$t>T$ and $\alpha\in\widehat{K}_r=\{\alpha\in\widehat{K}\,|\,\Re\alpha\geq r\}$.
Since for any $\alpha\in\widehat{K}$ there exists $\alpha'\in\widehat{K}_r$ such that
$|\alpha-\alpha'|<2r$ and therefore $|\cF_{\lambda,t}(\alpha)-\cF_{\lambda,t}(\alpha')|<\epsilon/6$,
one has
\begin{align*}
|\cF_{\lambda,t}(\alpha)-\cF_{\lambda,s}(\alpha)|
\leq |\cF_{\lambda,t}(\alpha)-\cF_{\lambda,t}(\alpha')|
&+|\cF_{\lambda,s}(\alpha)-\cF_{\lambda,s}(\alpha')|\\
&+|\cF_{\lambda,t}(\alpha')-\cF_{\lambda}(\alpha')|
+|\cF_{\lambda,s}(\alpha')-\cF_{\lambda}(\alpha')|<\varepsilon,
\end{align*}
for any $\alpha\in\widehat{K}$ and $s,t>T$. It follows that 
$\cF_{\lambda,t}$ converges uniformly in $K$ as $t\to\infty$,
the limiting function $\cF_\lambda$ being continuous.\hfill\qed      

Since the function $\rr\ni\alpha\mapsto\cF_\lambda(\i\alpha)$ is continuous, there
exists a unique Borel probability measure $\P_{\cR,\lambda}$ on $\rr$
such that
\[
\int_\rr\e^{\i\gamma\varsigma}\d\P_{\cR,\lambda}(\varsigma)
=\cF_\lambda(\i\gamma/\beta)
\]
for any $\gamma\in\rr$. An immediate consequence of the last lemma is:
\bep\label{thm-prop1}
For $0<|\lambda|<\lambda_0$, 
\[
\lim_{t\to\infty}\P_{\cR,\lambda,t}=\P_{\cR,\lambda}. 
\]
\eep

\bel\label{not-believe}
For $\gamma\in\rr$, 
$$
\cF(\i\gamma/\beta)=\lim_{\lambda\to0}\cF_\lambda(\i\gamma/\beta)
=\rho_\cS\left(\e^{-\i\gamma H_\cS}\right)\omega_\cS\left(\e^{\i\gamma H_\cS}\right).
$$
\eel

\proof For $s\in\rr$, set
\[
\cG_\lambda^{(1)}(s)
=\langle\widehat{\Omega}|\e^{-\i\beta s\widehat{L}_\lambda} \Omega_\lambda\rangle,
\qquad
\cG_\lambda^{(2)}(s)
=\langle\Omega_\lambda|\e^{\i\beta s\widehat{L}_\lambda}\Omega_\eta\rangle.
\]
Writing $\widehat{L}_\lambda=L_0+\lambda\pi(V)-J\pi(H_\cS)J$ and noticing that
$L_0+\lambda\pi(V)=L_\cR+\pi(H_\cS+\lambda V)-J\pi(H_\cS)J$ commutes with 
$J\pi(H_\cS)J$, we obtain
\[
\cG_\lambda^{(1)}(s)
=\langle\widehat{\Omega}|
\e^{-\i\beta sJ\pi(H_\cS)J}\e^{-\i\beta s(L_0+\lambda \pi(V))}\Omega_\lambda\rangle.
\]
Araki's perturbation theory (recall~\eqref{araki-1}--\eqref{araki-3}) implies that
the function
$$
\rr\ni s\mapsto\cG_\lambda^{(1)}(s)
=\langle\widehat{\Omega}|
\e^{-\i\beta sJ\pi(H_\cS)J}\e^{-\i\beta(s-\i/2)(L_0+\lambda \pi(V))}\Omega_0\rangle,
$$
has an analytic continuation to the strip $0<\Im s<\12$ which is bounded and continuous 
on its closure. Thus, for $\gamma\in\rr$, one has
\[
\cG_\lambda^{(1)}\left(\frac{\gamma}{\beta}+\12\,\i\right)
=\langle\widehat{\Omega}|\e^{(\frac{\beta}{2}-\i\gamma)J\pi(H_\cS)J}
\e^{-\i\gamma(L_0+\lambda\pi(V))}\Omega_0\rangle,
\]
and it immediately follows that
\beq\label{mama-mia}
\lim_{\lambda\to0}\cG_\lambda^{(1)}\left(\frac{\gamma}{\beta}+\12\,\i\right)
=\langle\widehat{\Omega}|J\pi(\e^{(\frac{\beta}{2}+\i\gamma)H_\cS})J\Omega_0\rangle
=Z^{-\12}\rho_\cS(\e^{-\i\gamma H_\cS}),
\eeq
where $Z=\tr(\e^{-\beta  H_\cS})$. Since 
$\Omega_\eta=Z^\12 J\pi(\e^{\frac\beta2H_\cS})J\Omega_0$, one has,
with the notation of the proof of Lemma~\ref{prop-instant-1},
\begin{align*}
\e^{\i\beta s\widehat{L}_\lambda}\Omega_\eta
=U_M^{\beta s}\e^{\i\beta sL_\cR}\Omega_\eta
&=U_M^{\beta s}\Omega_\eta
=Z^\12U_M^{\beta s}J\pi(\e^{\frac\beta2H_\cS})J\Omega_0
=Z^\12J\pi(\e^{\frac\beta2H_\cS})JU_M^{\beta s}\Omega_0\\
&=Z^\12J\pi(\e^{\frac\beta2H_\cS})J
\e^{\i\beta s\widehat{L}_\lambda}\e^{-\i\beta sL_\cR}\Omega_0
=Z^\12J\pi(\e^{\frac\beta2H_\cS})J\e^{\i\beta s\widehat{L}_\lambda}\Omega_0,
\end{align*}
and hence
\begin{align*}
\e^{\i\beta s\widehat{L}_\lambda}\Omega_\eta
&=Z^\12J\pi(\e^{\frac\beta2H_\cS})J\e^{\i\beta sJ\pi(H_\cS)J}\e^{\i\beta s(L_0+\lambda\pi(V))}\Omega_0\\
&=Z^\12\e^{\i\beta(s-\i/2)J\pi(H_\cS)J}\e^{\i\beta s(L_0+\lambda\pi(V))}\Omega_0.
\end{align*}
Araki's perturbation theory implies that the function
$$
\rr\ni s\mapsto\cG_\lambda^{(2)}(s)
=Z^\12\langle\Omega_\lambda|\e^{\i\beta(s-\i/2)J\pi(H_\cS)J}
\e^{\i\beta s(L_0+\lambda\pi(V))}\Omega_0\rangle
$$
also has 
an analytic continuation to the strip $0<\Im s<\12$ which is bounded and continuous on 
its closure. For $\gamma\in\rr$, one gets
\[
\cG_\lambda^{(2)}\left(\frac{\gamma}{\beta}+\12\,\i\right)
=Z^\12\langle\Omega_\lambda|J\pi(\e^{-\i\gamma H_\cS})J
\e^{\i\gamma(L_0+\lambda\pi(V))}\Omega_\lambda\rangle.
\]
Since 
\beq 
\lim_{\lambda\to0}\Omega_\lambda=\Omega_0,
\label{airport}
\eeq
we conclude that
\beq\label{tata-mia}
\lim_{\lambda\to0}\cG_\lambda^{(2)}\left(\frac{\gamma}{\beta}+\12\,\i\right)
=Z^\12\langle\Omega_0|J\pi(\e^{-\i\gamma H_\cS})J\Omega_0\rangle
=Z^\12\omega_\cS(\e^{\i\gamma H_\cS}).
\eeq
Finally, recall that for $s\in \rr$,
\[
{\cal F}_\lambda\left(\frac{1}{2} +\i s\right)=\frac{1}{\|\Omega_\lambda\|^2}{\cal G}_\lambda^{(1)}(s){\cal G}_\lambda^{(2)}(s).
\]
Analytic continuation of this relation to 
\[
s=\frac{\gamma}{\beta}+\12\,\i
\]
combined with~\eqref{mama-mia}--\eqref{tata-mia} gives the result. 
\hfill\qed

An immediate consequence of the last lemma (recall~\eqref{sale-2}) is 
\bep 
\[
\lim_{\lambda\to0}\P_{\cR,\lambda}=\P_\cS.
\]
\label{thm-prop2}
\eep
This completes the proof of Theorem~\ref{thm-instant-1}.
\appendix

\section{Summary of the algebraic framework}
\label{algebraic}

For the reader's convenience, we give here a simplified presentation of some well 
known constructions, restricting ourselves to the material we have been using throughout the 
paper. When possible, we give reference to precise parts of \cite{BR1,BR2}
where detailed proofs can be found.

We first recall some basic definitions.
\bed
\ben
\item A Banach algebra $\cA$ equipped with an involution $\ast$ is called 
$C^\ast$-algebra if
\[
\|A^\ast A\|=\|A\|^2
\]
for all $A\in\cA$. We always assume a $C^\ast$-algebra admits an identity denoted by $\one$.

\item A linear functional $\omega$ on $\cA$ is \emph{positive} if 
$\omega(A^\ast A)\geq0$ for all $A\in\cA$.
\item A positive linear functional on $\cA$ is a \emph{state} if $\omega(\one)=1$.
\item A state is said to be \emph{faithful} iff $\omega(A^\ast A)= 0$ implies $A=0$.
\een
\eed

\bep
A positive linear functional on a $C^\ast$-algebra is automatically continuous.
\eep

Let $\cA,\cB$ denote $C^\ast$-algebras.

\bed  A linear map $\phi:\cA\to\cB$ is called \emph{$\ast$-morphism} iff, for all 
$A,B\in\cA$,
\ben
\item $\phi(AB)=\phi(A)\phi(B)$,
\item $\phi(A^\ast)=\phi(A)^\ast$.
\een
If, furthermore, $\phi$ is bijective, it is called \emph{$\ast$-isomorphism}.
A $\ast$-isomorphism with $\cA=\cB$ is called a  \emph{$\ast$-automorphism}.
\eed

\bep
If $\phi$ is a $\ast$-morphism then $\norme{\phi(A)}\leq\norme{A}$. In particular, 
$\phi$ is continuous.
\eep

\bed
Let $\cH$ be a Hilbert space and $\cC\subset\cB(\cH)$. A vector $\Omega\in\cH$ is called \emph{cyclic} for $\cC$ iff  $\cC\Omega$ is dense in $\cH$,
and \emph{separating} iff $C\Omega=C'\Omega$ for some $C,C'\in\cC$ implies $C=C'$.
\eed

\bed A \emph{representation} of a $C^\ast$-algebra $\cA$ is a pair $(\cH,\pi)$ where 
$\cH$ is a complex Hilbert space and $\pi:\cA\to\cB(\cH)$ is a $\ast$-morphism.
A \emph{cyclic representation} of $\cA$ is a triple $(\cH,\pi,\Omega)$ such that 
$(\cH,\pi)$ is a representation of $\cA$ and the unit vector $\Omega\in\cH$ is cyclic 
for $\pi(\cA)$.
\eed

\subsection{Canonical cyclic representation}
\label{sec GNS}

Given a $C^\ast$-algebra $\cA$, it is always possible to find a cyclic representation 
(see \cite[Thm 2.3.16]{BR1}).
\bet[Gelfand-Naimark-Segal]\label{GNS}
Let $\omega$ be a state over the $C^\ast$-algebra $\cA$. There exists a cyclic 
representation $(\cH_\omega,\pi_\omega,\Omega_\omega)$ of $\cA$ such that, 
for all $A\in\cA$,
\[
\omega(A)=\langle\Omega_\omega|\pi_\omega(A)\Omega_\omega\rangle.
\]
Moreover, this representation is unique up to unitary equivalence.
\eet

\bed The triple $(\cH_\omega, \pi_\omega, \Omega_\omega)$  in the previous theorem  is called  \emph{ the canonical cyclic representation} or \emph{the GNS representation} 
of $\cA$ induced by $\omega$.
\eed

The representation $(\cH_\omega,\pi_\omega,\Omega_\omega)$ is constructed as follows.
Equip the vector space $\cA$ with the positive semi-definite sesquilinear form
\beq
\langle A|B\rangle=\omega(A^\ast B).
\label{preinner}
\eeq
By the inequality $\omega(B^\ast A^\ast AB)\leq\|A\|^2\omega(B^\ast B)$, the set
\[
\mathcal{I}_\omega=\{ A\in\cA\,|\,\omega(A^\ast A)=0\}
\]
is a closed (left) ideal and the quotient $\cA\slash\mathcal{I}_\omega$
equipped with the inner product induced by~\eqref{preinner} is a pre-Hilbert space, the
completion of which is $\cH_\omega$. For $A\in\cA$, define $\pi_\omega$ by
\[
\pi_\omega(A):X+\mathcal{I}_\omega\mapsto AX+\mathcal{I}_\omega.
\]
Clearly,\ the vector $\Omega_\omega=\one+\mathcal{I}_\omega$ is cyclic for $\pi_\omega({\cal A})$ and 
satisfies
\[
\omega(A)=\langle\Omega_\omega|\pi_\omega(A) \Omega_\omega\rangle
\]
for all $A\in\cA$.

\subsection{von Neumann Algebras}

\bed
A $C^\ast$-algebra $\fM\subset\cB(\cH)$ is called a \emph{von Neumann algebra}
if it is closed with respect to the strong topology of $\cB(\cH)$.
\eed

If $\cC\subset\cB(\cH)$, we denote by $\cC'$ its commutant, i.e., the set
$$
\cC'=\{A\in\cB(\cH)\,|\,[A,C]=0 \text{ for all } C\in\cC\}.
$$

Here we give some characterization of  von Neumann algebras.
\bet[von Neumann]\label{bicommutant}
Let $\fM\subset\cB(\cH)$ be a $C^\ast$-algebra.
The following conditions are equivalent:
\ben
\item $\fM$ is a von Neumann algebra.
\item $\fM=\fM''$.
\item $\fM$ is closed with respect to the weak topology of $\cB(\cH)$.
\een
\eet

\subsection{Normal states}
\label{sec normal states}

\bed
\ben
\item The \emph{$\sigma$-weak topology} on $\cB(\cH)$ is the locally convex topology induced
by the semi-norms
$$
A\mapsto\left|\sum_{n\in\nn}\langle\Phi_n|A\Psi_n\rangle\right|,
$$
where $\Phi_n,\Psi_n\in\cH$ are such that $\sum_{n}\|\Phi_n\|^2$ and
$\sum_{n}\|\Psi_n\|^2$ are finite.
\item A linear functional on the von Neumann algebra $\fM\subset\cB(\cH)$ is 
\emph{normal} if it is $\sigma$-weakly continuous.
\een
\eed

Several properties  can be used to characterize normal states of a von Neumann 
algebra, see~\cite[Thm 2.4.21]{BR1}. We recall the most useful one for this paper.
\bet
A state $\omega$ on von Neumann algebra $\fM\subset\cB(\cH)$ is normal iff there 
exists a density matrix $\rho_\omega$, i.e., a positive trace class operator on $\cH$ with
unit trace, such that $\omega(A)=\tr(\rho_\omega A)$ for all $A\in\fM$ .
\eet

Let $\omega$ be a state on the $C^\ast$-algebra $\cA$ and
$(\cH_\omega,\pi_\omega,\Omega_\omega)$ the induced GNS representation.
The von Neumann algebra $\pi_\omega(\cA)''\subset\cB(\cH_\omega)$ is called the 
\emph{enveloping von Neumann algebra.} $\pi_\omega(\cA)$ is $\sigma$-weakly
dense in $\pi_\omega(\cA)''$ and the state $\omega$ has a unique normal extension
to $\pi_\omega(\cA)''$ given by
\beq
\widehat{\omega}(A)=\langle\Omega_\omega|A\Omega_\omega\rangle.
\label{hatomega}
\eeq

\bed\label{omeganormdef}
A state $\eta$ on $\cA$ is $\omega$-normal if it extends to a normal state on
$\pi_\omega(\cA)''$, i.e., if there exists a density matrix $\rho_\eta$ on $\cH_\omega$
such that $\eta(A)=\tr(\rho_\eta\pi_\omega(A))$.
\eed

\subsection{KMS states}
\label{sec-KMS}

\bed
A $C^\ast$ (resp. $W^\ast$) dynamical system is pair $ (\cA,\tau)$ where $\cA$ is a $C^*$-algebra (resp. a von Neumann algebra) and $\tau$ is a strongly 
(resp. $\sigma$-weakly) continuous one-parameter group of $\ast$-automorphisms of $\cA$.
A state $\omega$ on $\cA$ is $\tau$-invariant if $\omega\circ\tau^t=\omega$ for all
$t\in\rr$.
\eed

\bed
Let $(\cA,\tau)$ be a $C^\ast$ (resp. $W^\ast$) dynamical system. $A\in\cA$ is called 
\emph{analytic} for $\tau$ if there exists a function $f:\cc\to\cA$ such that
\ben 
\item $f(t)=\tau^t (A)$ for $t\in\rr$,
\item for any state (resp. any normal state) $\eta$ on $\cA$, the function 
$z\mapsto\eta(f(z))$ is entire analytic.
\een 
The set of analytic elements for $\tau$ is denoted by $\cA_\tau$.
\eed

\bet The set $\cA_\tau$ is a dense (resp. $\sigma$-weakly dense) $\ast$-subalgebra of 
$\cA$.
\eet

\bed
Let $(\cA, \tau)$ be a $C^*$ (resp. $W^\ast$) dynamical system and $\beta\in\rr$. 
A state (resp. a normal state) $\omega$ on $\cA$ is said to be a $(\tau,\beta)$-KMS
state iff 
\[
\omega(A\tau^{\i\beta}(B))=\omega(BA) 
\]
for all $A,B\in\cA_\tau$.
\eed

A $(\tau,\beta)$-KMS state describes a thermal equilibrium state at 
inverse temperature $\beta$. In particular, it is $\tau$-invariant 
(\cite[Prop. 5.3.3]{BR2}). 
If $\cA$ is finite dimensional and $\tau^t(A)=\e^{\i tH}A\e^{-\i tH}$ for some self-adjoint Hamiltonian $H\in\cA$, then the state
$$
\omega(A)=\frac{\tr(\e^{-\beta H}A)}{\tr(\e^{-\beta H})}
$$
is the unique $(\tau,\beta)$-KMS state.

We note also that if $\beta,\tilde{\beta}\in\rr\setminus\{0\}$ and $\omega$ is 
$(\tau,\beta)$-KMS, then it is also $(\tilde{\tau},\tilde{\beta})$-KMS for the
dynamics $\tilde{\tau}^t=\tau^{t\beta/\tilde{\beta}}$.

\bep
Let $(\cA,\tau)$ be a $C^\ast$-dynamical system and $\omega$ a $(\tau,\beta)$-KMS 
state for some $\beta\in\rr$. Let $(\cH_\omega,\pi_\omega,\Omega_\omega)$ 
be the induced GNS representation.
\ben
\item The cyclic vector $\Omega_\omega$ is separating for the enveloping von Neumann 
algebra $\pi_\omega(\cA)''$.
\item The state $\widehat\omega$ (the normal extension~\eqref{hatomega} of $\omega$ to 
$\pi_\omega(\cA)''$) is faithful.
\item If $\beta\not=0$, there exists a unique $W^\ast$-dynamical system 
$(\widehat{\tau},\pi_\omega(\cA)'')$ such that $\widehat{\tau}^t(\pi_\omega(A))=\pi_\omega(\tau^t(A))$ for all $t\in\rr$ and
$A\in\cA$.
\item $\widehat{\omega}$ is $(\widehat{\tau},\beta)$-KMS.
\een
\eep

\subsection{Modular theory}
\label{modular operator app}

Given a von Neumann algebra $\fM\subset\cB(\cH)$ and a vector $\Omega\in\cH$ which is 
separating for
$\fM$, the map  
\[S(A\Omega)=A^\ast\Omega\]
defines an anti-linear operator on the 
subspace $\fM\Omega$. If $\Omega$ is cyclic for $\fM$, then this operator is densely defined.
\bep {\rm \cite[Prop. 2.5.9]{BR1}}
If $\Omega$ is cyclic and separating for $\fM$, then $S$ has a closed extension
$\bar S$ with dense domain $\Dom(\bar S)\supset\fM\Omega$.
\eep

Since $\bar S$ is a closed operator it admits a unique polar decomposition given by
\[
\bar S=J\Delta^\12,
\]
where $J$ is anti-unitary and  $\Delta=\bar S^\ast \bar S$ is positive. By definition, 
one has
\beq
J\Delta^\12A\Omega=A^\ast\Omega
\eeq
for all $A\in\fM$. Let $\cP\subset\cH$ denote the closure of the set 
$\{AJAJ\Omega\,|\,A\in\fM\}$.

\bed\label{defmodular}
$\Delta$ is the modular operator, $J$ the modular conjugation, and $\cP$ the natural
cone of the pair $(\fM,\Omega)$.
\eed

Some basic properties that follow easily from the definition are
$J^2=1$, $J\Omega=\Omega$, $\Delta\Omega=\Omega$, $J\Delta^{\12}=\Delta^{-\12}J$
(see \cite[Prop. 2.5.11]{BR1}). Much deeper is the following:

\bet[Tomita-Takesaki]
\label{TT}
Let $\Delta$ and $J$ be the modular operator and modular conjugation of $(\fM,\Omega)$.
Then,
\[
J\fM J=\fM',
\]
and for all $t\in\rr$,
\[
\Delta^{\i t}\fM\Delta^{-\i t}=\fM,
\]
$$
\Delta^{\i t}\cP=\cP.
$$
\eet
\bec
The modular operator of $(\fM,\Omega)$ defines a $W^\ast$-dynamical system 
$(\fM,\sigma)$ given by 
\[\sigma^t(A)=\Delta^{\i t}A\Delta^{-\i t}.\]
Moreover, 
the state
$$
A\mapsto\frac{\langle\Omega|A\Omega\rangle}{\|\Omega\|^2}
$$
is KMS for $\sigma$ at inverse temperature $\beta=-1$.
\eec
\bed
$\sigma$ is the \emph {modular group} of the pair $(\fM,\Omega)$.
\eed

\bet[Takesaki]\label{TakesakiThm}
The modular group $\sigma$ is the unique dynamics on $\fM$ for which the state
$\omega$ is KMS at inverse temperature $\beta=-1$.
\eet

\bed
A state $\omega$ on the $C^\ast$-algebra $\cA$ is called \emph{modular} if the cyclic
vector $\Omega_\omega$ of the induced GNS representation is separating for the
enveloping von Neumann algebra $\pi_\omega(\cA)''$. In this case, we denote by
$\Delta_\omega$, $J_\omega$, $\cP_\omega$ and $\sigma_\omega$ the modular operator, 
the modular conjugation, the natural cone and the modular group of the pair $(\pi_\omega(\cA)'',\Omega_\omega)$.
\eed

\bep\label{PropJ}{\rm\cite[Proposition 2.5.30, Theorem 2.5.31]{BR1}}
Let $\omega$ be a modular state on the $C^\ast$-algebra $\cA$. For any
$\omega$-normal state $\nu$ on $\cA$ there exists a unique unit vector
$\Omega_\nu\in\cP_\omega$ such that 
$\nu(A)=\langle\Omega_\nu|\pi_\omega(A)\Omega_\nu\rangle$ for all $A\in\cA$.
Moreover, $\Omega_\nu$ is separating for $\pi_\omega(\cA)''$ iff it is cyclic.
In that case, the modular conjugation and the natural cone of the pair
$(\pi_\omega(\cA)'',\Omega_\nu)$ satisfy $J_\nu=J_\omega$ and $\cP_\nu=\cP_\omega$.
\eep

\bed
$\Omega_\nu$ is the \emph{standard vector representative} of the state $\nu$.
\eed

We  now  state an important consequence of Takesaki's theorem (see \cite[Thm. 5.3.10]{BR2}).

\bep\label{identify dynamics}
Let $\omega$ be a modular state on the $C^\ast$-algebra $\cA$ and 
$\beta\in\rr\setminus\{0\}$. Then $\widehat{\tau}^t=\sigma_\omega^{-t/\beta}$
defines the unique $W^\ast$-dynamical system on $\pi_\omega(\cA)''$ such that the 
normal extension $\widehat{\omega}$ is $(\widehat{\tau},\beta)$-KMS.
\eep

By the above proposition, given a $(\tau,\beta)$-KMS state $\omega$, 
the relation 
\beq
\pi_\omega(\tau^t(A))=\sigma_\omega^{-t/\beta}(\pi_\omega(A))
\label{tausigma}
\eeq
identifies $\tau$ with the modular dynamics $\sigma_\omega$.
\subsection{The standard Liouvillean}
\label{GNS liouvillean app}

Concerning the implementation of $C^\ast$-dynamical systems by unitary groups
in GNS representations, one has the following general result 
(see~\cite[Corollary 2.5.32]{BR1} and \cite[Theorem 4.43]{Pi}).

\bep
Let $\omega$ be a modular state on the $C^\ast$-algebra $\cA$ and denote by
$(\cH_\omega,\pi_\omega,\Omega_\omega)$ the induced GNS representation. For any
strongly continuous one-parameter group $\tau$ of $\ast$-automorphisms of $\cA$
there exists a unique self-adjoint operator $L$ on $\cH_\omega$ such that,
for all $t\in\rr$,
\ben
\item
$$
\e^{\i tL}\cP_\omega=\cP_\omega.
$$
\item
$$
\e^{\i tL}\pi_\omega(A)\e^{-\i tL}=\pi_\omega(\tau^t(A))
$$
for all $A\in\cA$.
\een
\eep

\bed
The operator $L$ is the \emph{standard Liouvillean} of the triple $(\cA,\tau,\omega)$.
\eed

The standard Liouvillean $L$ also satisfies $[J_\omega,\e^{\i tL}]=0$, from
which one deduces
$$
\e^{\i tL}\pi_\omega(\cA)'\e^{-\i tL}\subset\pi_\omega(\cA)'.
$$

If $\omega$ is a $(\tau,\beta)$-KMS state, the identification~\eqref{tausigma} of 
$\tau$ with the modular group $\sigma_\omega$ given by Takesaki's 
theorem~\ref{TakesakiThm} translates as 
\beq
L=-\frac1\beta\log\Delta_\omega.
\label{LDeltaForm}
\eeq

\subsection{Relative modular operator}

\label{sec- relative modular}

Let $\fM\subset\cB(\cH)$ be a von Neumann algebra and $\Psi,\Phi\in\cH$.
If $\Phi$ is separating for $\fM$, then 
$S_{\Psi|\Phi}(A\Phi)=A^\ast\Psi$ defines an anti-linear operator
on the subspace $\fM\Phi$.

\bep[\cite{Ar1,Ar2}]
If $\Phi$ is cyclic and separating for $\fM$, then $S_{\Psi|\Phi}$ has a closed
extension $\bar S_{\Psi|\Phi}$ with a dense domain 
$\Dom(\bar S_{\Psi|\Phi})\supset\fM\Phi$. If $J$ is the
modular conjugation of the pair $(\fM,\Phi)$, then
$$
\bar S_{\Psi|\Phi}=J\Delta_{\Psi|\Phi}^\12,
$$
where $\Delta_{\Psi|\Phi}=\bar S_{\Psi|\Phi}^\ast \bar S_{\Psi|\Phi}$ is positive.
\eep

\bed
Let $\omega$ be a modular state on the $C^\ast$-algebra $\cA$. For any $\omega$-normal
state $\nu$ on $\cA$, we define the relative modular operator of $\nu$ w.r.t.\;$\omega$
by
$$
\Delta_{\nu|\omega}=\Delta_{\Omega_\nu|\Omega_\omega},
$$
where $\Omega_\omega,\Omega_\nu\in\cP_\omega$ are the standard vector representatives
of $\omega$ and $\nu$.
\eed

\end{document}